\documentclass[manuscript]{aastex}

\begin{document}
\pagenumbering{arabic}

\title{GLOBULAR CLUSTERS IN DWARF AND GIANT GALAXIES}

\author{Sidney van den Bergh}
\affil{Dominion Astrophysical Observatory, Herzberg Institute of Astrophysics, National Research Council of Canada, 5071 West Saanich Road, Victoria, BC, V9E 2E7, Canada}
\email{sidney.vandenbergh@nrc.gc.ca}

\begin{abstract}

The luminosity distribution of globular clusters shows a dramatic
dependence on parent galaxy luminosity. Dwarf galaxies contain 
far more faint globulars than do luminous galaxies. This  difference 
is significant at the 99.7\% level. On the other hand the luminosity
distribution of globular clusters in dwarf galaxies does not appear
to depend strongly on their host's  morphological type. The dichotomy of globular cluster masses occurs at a host galaxy mass of $\sim 4 \times 10^{8}$ M$_\odot$, which is almost two orders of magnitude lower than the onset of the dichotomy in globular color characteristics at $\sim 3 \times 10^{10}$ M$_\odot$ that was recently noted by Forbes. 
 
\end{abstract}

\section{INTRODUCTION}

  According to current ideas on galaxy formation dwarf galaxies and giant
galaxies follow quite different evolutionary tracks. One might therefore 
expect these two types of galaxies to also have experienced very different 
histories of star and globular cluster formation. The first indication for
such differences  (van den Bergh 1975) was provided by the observation that 
the globulars associated with giant galaxies are systematically more metal-
rich than those hosted by dwarf galaxies. It is the purpose of the present 
investigation to see if there are also systematic differences between the 
luminosities (masses) of the globular clusters associated with giant and dwarf
galaxies. The main difficulty with such an investigation is that it is not easy 
to find galaxy samples for which the globular cluster data are complete down
to sufficiently faint magnitude limits. 

\section{DATA ON GLOBULAR CLUSTERS}

A recent Hubble Space Telescope snapshot survey of nearby  dwarf galaxies 
with ${\it D} < 4.0$ Mpc by Sharina, Puzia \& Marakov (2005) [hereinafter referred
to as SPM] has provided a compilation of information on  clusters  which is 
essentially complete down to ${\it M}_{v} \sim -5.0$ . Dwarf galaxies are mostly low in metals.  One therefore expects the clusters within them to generally 
exhibit  low reddening values. Most of the red clusters with ({\it V - I}) $>$ 0.70 in the SPM catalog are therefore probably globular clusters, rather than reddened young blue clusters.  

A listing of such red clusters in galaxies with {\it D} $<$ 4.0 Mpc is given in Table 1. For more distant host galaxies the SPM data suffer from increasing incompleteness for faint clusters. Since most of the galaxies contained in the SPM  catalog are situated in small clusters they are located in  environments similar to that of our own Local Group. (Hubble 1936, van den Bergh 2000). It therefore seems reasonable to combine the cluster data collected by SPM with those on globular clusters in the Local Group. For those 
globular  clusters in Local Group dwarf galaxies that are situated within
150 kpc the data on the globular clusters luminosity distribution are probably also complete down to ${\it M}_{v} \sim -5.0$. The data  (Mackey \& van den Bergh 2005) for the globular clusters in the LMC, the SMC, the Fornax dwarf and the Sagittarius  dwarf are listed in Table 1.  Additional information, which may not be quite as complete at faint magnitudes, is available for for the globular clusters in the more distant Local Group dwarfs NGC 205 ({\it D} = 760 kpc),  NGC 6822  ({\it D} = 500 kpc), NGC 185 ({\it D} = 660 kpc), NGC 147 ({\it D} = 660 kpc) and the Wolf-Lundmark-Melotte system ({\it D} = 925 kpc). Data on the  luminosities of the globular clusters in these galaxies was taken from Hodge (1974), Hodge (1976), Ables \& Ables (1977),  Harris \& Racine (1979), Da Costa \& Mould (1988),  Harris (1991), Wyder, Hodge \& Zucker (2000), Strader, Brodie \& Huchra  (2003) and Da Costa (2003). The uncertainties and ambiguities in the identifications of the faintest globular clusters in distant Local Group galaxies are well summarized in the appendix to the paper by Ford, Jacoby \& Jenner (1977). Additional evidence for the possible incompleteness of the data sample on faint globular clusters in the most distant Local Group  galaxies is provided by the very recent discovery of two additional globular cluster suspects in the outer regions of NGC 6822  (Hwang et al. 2005). To remind the reader of Table 1 of this possible incompleteness  entries for the parent galaxies of  Local Group galaxies with {\it D} $>$ 150 kpc have been marked by an asterisk.

\section{COMPARISON OF THE GLOBULAR CLUSTERS IN DWARF AND IN GIANT GALAXIES}

 Available data  (Harris 1991, Di Criscienzo et al. 2006) suggest that most giant galaxies have similar log-normal cluster luminosity distributions that peak at ${\it M}_{v} \sim -7.6$.  The question that we wish to ask is:  Do the globular cluster systems associated with dwarf galaxies also have a similar luminosity distribution? In attempting to answer this question the luminosity distribution of Galactic globulars associated with the main body of the Galaxy, i. e. those having ${\it R}_{gc} <$ 15 kpc (van den Bergh \& Mackey 2004, Mackey \& van den Bergh 2005), for which data are most complete and reliable, will be used as the template for the cluster systems hosted by giant galaxies. This template can then be compared to the luminosity distributions of the globular clusters hosted by various classes of fainter parent galaxies. 
  
   Using the listing of van den Bergh \& Mackey (2004) for the combined
data on the 16 globulars in the LMC (${\it M}_{v}$ = -18.5) and the single globular
cluster in the SMC $({\it M}_{v} = -17.1)$ one finds that these objects have a mean
absolute magnitude $<{\it M}v> = -7.10$ $\pm$ 0.25. A Kolmogorov-Smirnov test
shows no significant difference between the luminosity distributions of the 
17 globular clusters in the Magellanic Clouds and the 110 globulars with $\it {R}_{gc} <$ 15 kpc that are hosted by the Galaxy. It is therefore tentatively concluded
that the luminosity distribution of the globulars in the Magellanic Clouds is 
similar to that in typical giant galaxies.

Table 2 and Figure 1  show a comparison between the luminosity distributions of the globular clusters associated with the main body of the Galaxy and that
of the nearby globulars associated with dwarf galaxies having {\it M}$_{v} > -17.0$. 
The Figure shows that the Galactic globular clusters have an approximately
log-normal luminosity distribution with a maximum between {\it M}$_{v} = -7.0$
and ${\it M}_{v} = -7.5$, whereas the globulars associated with dwarf galaxies appear
to have a luminosity distribution that exhibits a monotonic increase down to
the completeness limit of the data at ${\it M}_{v} \sim -5$. A Kolmogorov-Smirnov
 test on the data in Table 2 shows that there is only a 0.9\% probability that
the data for the Galactic clusters and for the globulars associated with
all nearby dwarf galaxies were drawn from the same parent population. A
similar test shows that the luminosity distributions of the globulars hosted 
by  ``bright dwarfs'' with $-17.0 > {\it M}_{v} > -14.5$ differs from that of the ``faint dwarfs'' with $-14.5 > {\it M}_{v} > -11.0$ at the 5\% significance level. The observed 
difference is in the sense that the luminosity distribution of globular 
clusters in faint dwarfs is more heavily weighted towards faint clusters
than is that of the globular clusters hosted by giant galaxies. The data in
Table 1 show that this effect is particularly pronounced for the cluster 
systems hosted by very faint parent systems having ${\it M}_{v} > -14.5$. A K-S
test shows that there is only a 0.3\% probability that the 40 globulars in
such faint dwarfs were drawn from the same parent population as are
the 110 globular clusters associated with the main body of the Galaxy. 
The luminosity distribution of the clusters hosted by dwarfs  fainter
than ${\it M}_{v} = -14.5$ peaks at ${\it M}_{v} \sim -5.25$, i. e. just above the completeness
limit of the data at ${\it M}_{v}  \sim -5.0$. This strongly suggests that the difference
between the luminosity distributions of globular cluster systems hosted
by dwarf and giant galaxies would be  even more striking if cluster
data could be extended to fainter completeness limits.

\section{DISCUSSION}

It is interesting to ask if the characteristics of the globular cluster
systems hosted by dwarf galaxies depend not only on host galaxy
luminosity, but also on the morphological types of these hosts. In
first approximation the dwarfs in Table 1 may be divided into two
broad morphological types: (1) Early-type galaxies with de Vaucouleurs
type ${\it T} < 0$ , and (2) late-type galaxies with ${\it  T} > 0$. A K-S test shows
no significant difference between the luminosity distributions of 
the 44 globuar clusters in early type galaxies and that of the 13 globulars
hosted by late-type galaxies. It should, however, be emphasized that
this conclusion is based on a relatively small data sample. It would
clearly be of interest to extend this data base by carrying out HST
snapshot surveys of larger samples of  galaxies with distances 
smaller than 4 Mpc. Nevertheless the data that are already in
hand clearly suggest that: (1) The luminosity distribution of  globular
clusters in dwarf galaxies differs dramatically from that in giants,
and (2) there is no evidence that the luminosity distribution of the
globular clusters in dwarf galaxies depends on the morphological
types of their hosts. Observations by Chandar, Whitmore \& Lee (2004)
suggest that M101, which appears to have a power-law luminosity
distribution of globular clusters down to ${\it M}_{v} \sim -6$, may be an
exception to the rule that giant galaxies contain few faint globulars.

The observation that faint globular clusters are common in dwarf
galaxies, but rare in giants, might be accounted for by the fact that disk 
and bulge shocks will destroy clusters more efficiently in giants than
in dwarfs. However, it is not yet possible to exclude the alternative
possibility that real differences in star and cluster forming conditions
in dwarf and giant galaxies were responsible for the observed
differences between the luminosity distributions of globular clusters 
hosted by dwarf and giant galaxies. For example Dekel \& Birnboim 
(2004) have recently suggested that halos below a certain shock-heating
mass will build star and cluster forming disks from cold streams, whereas
the clusters in more mossive halos might  form from in hot flows.  For an alternative suggestion about the early history of globular cluster formation the reader is referred to van den Bergh (2001).

 Recently Forbes (2005) has drawn attention to the fact that globular cluster systems with host masses greater than 3 $\times 10^{10}$ M$_\odot$ tend to have globular clusters with bimodal color distributions, whereas the clusters surrounding less massive galaxies are embedded in unimodal blue globular cluster systems. The dichotomy in the globular cluster luminosities occurs at host galaxy masses of $\sim 4 \times 10^{8}$ M$_\odot$. This difference of almost two orders of magnitude in host galaxy masses shows that the effects discussed in the present paper, and those to which Forbes has drawn attention, are probably due to different physical causes. In particular it is noted that the unimodel blue LMC globular cluster system belongs to Forbes' low mass parent class, whereas the luminosity distribution of the LMC's globulars places it among the subset of galaxy hosts that have high galaxy luminosities and masses.

   I am indebted to Narae Hwang for permission to include the results on two new globular clusters in NGC 6822 in the present paper. I also thank Avishai Dekel, Oleg Gnedin, and Dougal Mackey for helpful exchanges of e-mail.

  

         

\begin{deluxetable}{llllr}
\tablewidth{0pt}        
\tablecaption{Globular clusters in galaxies with ${\it D} < 4.0$ Mpc that are fainter than ${\it M}_{v} = -18.9$}

\tablehead{\colhead{Galaxy} & \colhead{{\it T}}  & \colhead{${\it M}_{v}$}  & \colhead{Cluster}  & \colhead{${\it M}_{v}$}}

\startdata

LMC   &   9  &   -18.5   &   N 1466   &   -7.26\\
      &      &           &   N 1754   &   -7.09\\
      &      &           &   N 1786   &   -7.70\\
      &      &           &   N 1835   &    -8.30\\
      &      &           &   N 1841   &   -6.82\\
      &      &           &  N 1898    &   -7.49\\
      &       &          &  N 1916     &  -8.24\\
      &      &           &   N 1928    & -6.06:\\
      &       &          &   N 1939   &   -6.85:\\
       &      &           &   N 2005    & -7.40\\
       &      &          &   N 2019     &  -7.75\\
       &      &          &  N 2210      & -7.51\\
       &      &          &  N 2257      &  -7.25\\
       &      &          &   Hodge 11   &  -7.45\\
       &      &          & Retic        &  -5.22\\
       &      &          &   ESO121     &  -4.37\\
SMC    & 9    &  -17.1    &  N 121      & -7.89\\
NGC 205* & -5 &  -16.4    &  Hu I       &  -7.62\\
      &       &           &  Hu II       &  -7.82\\
      &        &           &  Hu IV      &  -6.02\\
      &        &           &  Hu VI      &  -6.62\\
      &        &           &  Hu VII      &  -6.52\\
      &        &            &  Hu VIII    &  -7.92\\
NGC 6822* & 10 &   -16.0    &   Hu VII     &   -8.5\\
       &  &                 & SC 1         &   -7.26\\
       &  &                 &  SC 2        &   -5.6\\

NGC 185*  &   -5     & -15.6   &  I       &   -6.31\\
          &          &          & II      & -4.98\\
          &          &          &  III    &    -7.91\\
          &          &          &  IV      &   ...\\
          &           &       &   V       &   -7.97\\
          &           &        &  VI      &   ... \\

NGC 147* &  -5       &-15.1    & No. 1     & -6.97\\
         &           &         & No. 2     & -6.25\\
         &           &         & No. 3     &  -7.65\\
         &           &         & No. 4     & -3.53\\

WLM*    & 10         & -14.4   & ...  &   -8.33\\

Ho. IX  & 10        &   -13.8    & 3-1565    &   -5.31\\
        &           &            & 3-1932   & -6.61\\
        &           &             & 3-2373   &  -6.04\\

Sagit.  & -5       &  -13.8      &   N 4147  &   -6.16\\
         &         &              &  N 6715   &  -10.01 \tablenotemark{a}\\
         &          &             &  Ter. 7   &  -5.05\\
         &          &             &  Arp 2     &  -5.29\\
         &          &             &  Ter. 8   &  -5.05\\
         &          &             &  Pal. 12  &   -4.48\\

DDO 53  & 10       &  -13.74      &   3-1120  &   -5.88\\

KDG 61  & -1       &  -13.58      &  3-1325   &  -7.55\\

Fornax  &-5        &  -13.1       &  No. 1    &  -5.32\\
        &          &              &  No. 2    &  -7.03\\
        &          &              & NGC 1049  & -7.66\\
        &          &              &  No. 4    &   -6.83\\
        &          &              &  No. 5    &   -6.82\\

KDG 63  &-3        & -12.82       &   3-1168  &  -7.09\\

DDO 78  & -3       & -12.75        &   1-167   & -7.23\\
        &           &              &  3-1082   &  -8.81\\

DDO 113 & 10       & -12.67       &  2-579     &   -5.60\\
         &         &               &   4-690   &  -5.27\\

KK 211  & -5      &   -12.58       &   3-917   &   -6.86\\
        &          &               &  3-149    &   -7.82\\

KK 27  & -3        & -12.32        &   4-721  &  -6.36\\

KK 77  &  -3       &  -12.21       &   4-939   & -5.01\\
        &          &               &   4-1162  & -5.37\\
        &          &               &  4-1165   &  -5.69\\

KK221  &  -3       & -11.96        & 2-608     &  -8.04\\
       &           &               &  2-883    &  -7.07\\
       &           &               & 2-966     & -9.80\\
       &           &               & 2-1090    & -7.77\\
       &           &               & 3-1062    & -6.10\\
UA 438 & 10        &  -11.94       &  3-2004    & -8.67\\
       &            &              & 3-3325     &  -5.96\\

BK 6N  & -3       &   -11.93       &   2-524   &   -5.40\\
        &          &               &   4-789   &   -5.60\\

E 540  &  -3      & -11.84         &  4-1183   &  -5.37\\

E 294 &  -3       & -11.40         & 3-1104    & -5.32\\

KDG 73 & 10       & -11.31         & 2-378    &  -5.75\\

\tablenotetext{a}{ Probably a galaxy nucleus}
\enddata
\end{deluxetable}

\begin{deluxetable}{cccc}
\tablewidth{0cm}

\tablecaption{Luminosity distribution of globular clusters.}

\tablehead{
\colhead{{\it M}$_{v}$} & \colhead{Galaxy} & \colhead{Dwarf galaxies} &  \colhead{Dwarf galaxies}\\

                        & \colhead{{\it R}$_{gc} <$ 15 kpc} & \colhead{{\it M}$_v > -17.0$} 
& \colhead{{\it M}$_v > -14.5$}}
\startdata
-10.25  &  1\tablenotemark{a} & 1\tablenotemark{a}  &   1\tablenotemark{a}\\
-9.75   &  1                  &  1                  &  1  \\    
-9.25   &   7                 &  0                  &  0     \\
-8.75   &   8                 &     3               &  2     \\
-8.25   &   13                 &     2              &  2     \\
-7.75   &   18                 &     10             &  4     \\
-7.25   &   19                 &     5              &  4     \\
-6.75   &   19                 &     7              &  4     \\
-6.25   &   9                 &    7                &  4     \\
-5.75   &   5                 &     7               &  6     \\
-5.25   &  2                 &     11               &  11     \\
-4.75   &   3                 &     1               &  0     \\
-4.25   &   2                 &     1               &  1     \\   
-3.75   &   3                 &     1               &  0     \\
total   &   110               &     57              &  40\\

\tablenotetext{a}{dwarf galaxy core?}
\enddata
\end{deluxetable}

\begin{figure}[p]
\caption{Luminosity distributions of globular clusters. The figure shows that the globular clusters associated with giant Galaxy, which is a giant system,  have an approximately log-normal distribution, whereas those hosted by dwarf galaxies fainter than ${\it M}_{v} = -17.0$ appear to have a luminosity distribution that rises up to the completeness limit of the data at ${\it M}_v \sim -5$.}
\end{figure}

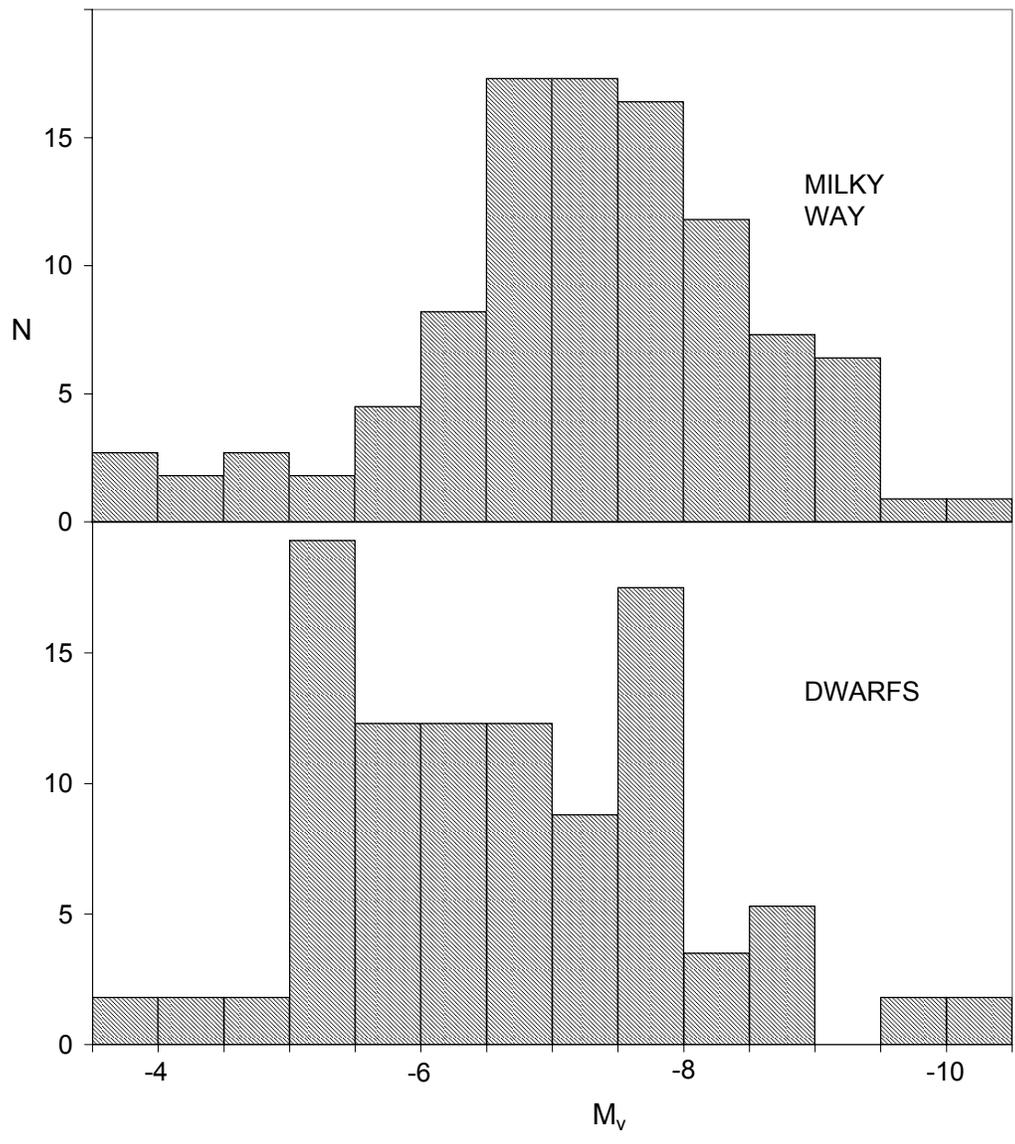
\end{document}